\documentstyle[aps,epsf,multicol,psfig]{revtex}
\begin{document}
\draft
\title{Elasticity theory connection rules for epitaxial interfaces}
\author{Corey W. Bettenhausen, Wade C. Bowie, and Michael R. Geller}
\address{Department of Physics and Astronomy, University of Georgia, Athens, 
Georgia 30602-2451}
\date{March 2, 2001}
\maketitle

\begin{abstract}
Elasticity theory provides an accurate description of the long-wavelength
vibrational dynamics of homogeneous crystalline solids, and with supplemental
boundary conditions on the displacement field can also be applied to abrupt 
heterojunctions and interfaces. The conventional interface boundary 
conditions, or connection rules, require that the displacement field and its 
associated stress field be continuous through the interface. We argue, 
however, that these boundary conditions are generally incorrect for epitaxial 
interfaces, and we give the general procedure for deriving the correct 
conditions, which depend essentially on the detailed microscopic structure of 
the interface. As a simple application of our theory we analyze in detail a 
one-dimensional model of an inhomogeneous crystal, a chain of harmonic 
oscillators with an abrupt change in mass and spring stiffness parameters. Our
results have implications for phonon dynamics in nanostructures such as 
superlattices and nanoparticles, as well as for the thermal boundary 
resistance at epitaxial interfaces.
\end{abstract}

\vskip 0.05in
\begin{multicols}{2}

\section{introduction}
\label{introduction section}

Continuum elasticity theory was developed in the 18th and 19th 
centuries---prior to the general acceptance of the atomic view of matter---to 
describe the mechanics of  elastic solids\cite{Love}. Modern 
applications of elasticity theory abound throughout science and engineering, 
from providing a long-wavelength description of the dynamics of crystalline 
lattices, to the inversion of seismological data to image the 
three-dimensional structure of the earth's interior.

The fundamental degree-of-freedom in a nonpolar elastic medium is the
displacement field ${\bf u}({\bf r})$, the deviation of the medium at point 
${\bf r}$ from its position in mechanical equilibrium. When applied to 
composite media consisting of layers or regions of different materials, 
characterized by different elastic parameters, a question naturally arises: 
What boundary conditions should be imposed on the displacement field 
at the interfaces?

\begin{figure}
\centerline{\psfig{file=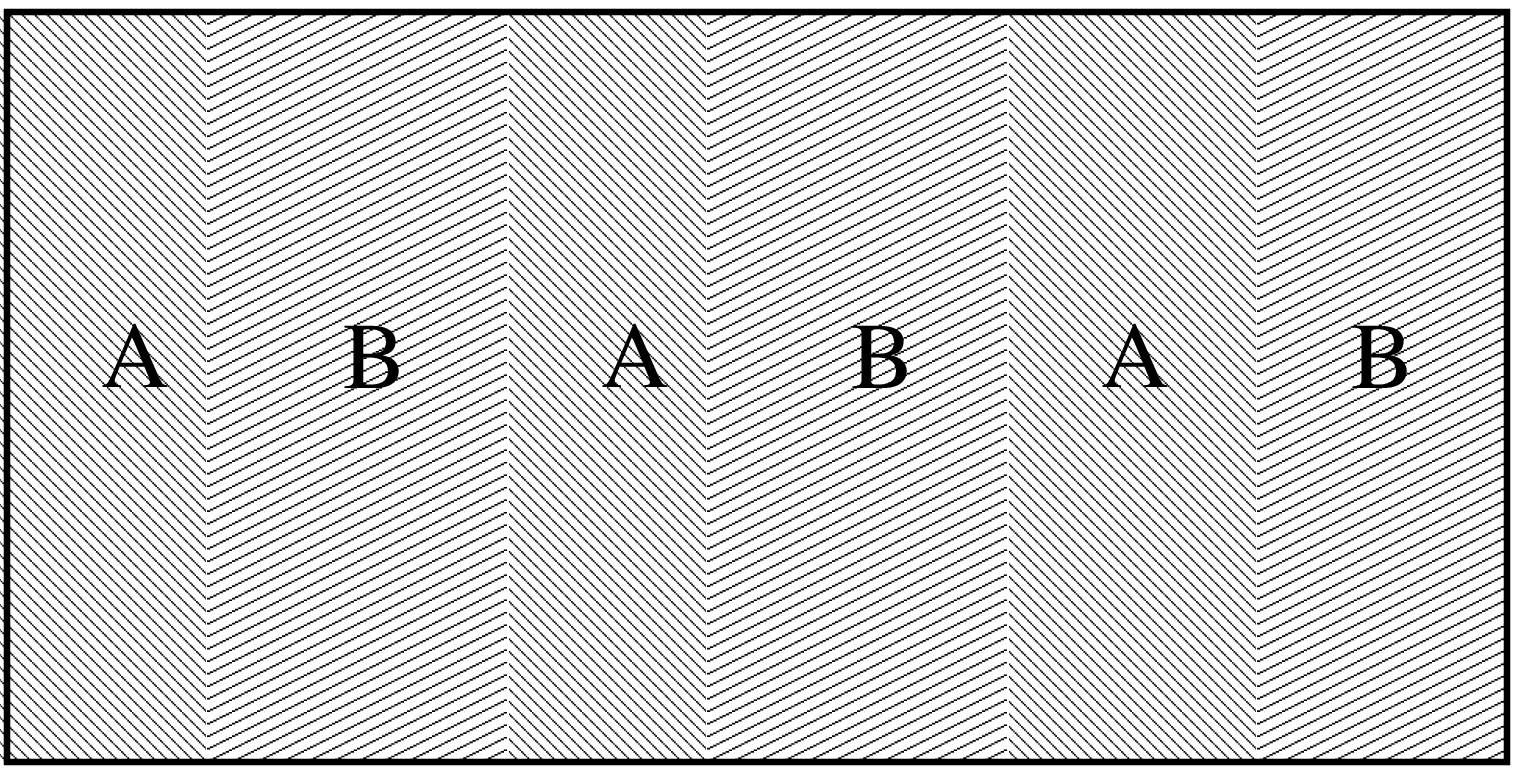,width=2.5in}}
\vspace{0.1in}\setlength{\columnwidth}{3.2in}
\centerline{\caption{
Superlattice consisting of layers of dissimilar elastic media, A and B.
\label{superlattice figure}}}
\end{figure}

An example of such a composite system is shown schematically in 
Fig.~\ref{superlattice figure}. Alternating layers of type A and type B 
materials, each characterized by different elastic constants and mass 
densities, are separated by abrupt interfaces. Within each region the
displacement field satisfies an appropriate equation of motion. For an
isotropic continuum with mass density $\rho$, the field equation is
\begin{equation}
\rho \, \partial_t^2 {\bf u} = v_{\rm l}^2 \mbox{\boldmath $\nabla$}
(\mbox{\boldmath $\nabla$} \cdot {\bf u}) - v_{\rm t}^2
\mbox{\boldmath $\nabla$} \times \mbox{\boldmath $\nabla$} \times {\bf u}, 
\label{elasticity equation}
\end{equation}
where $v_{\rm l} \equiv \sqrt{(\lambda + 2 \mu)/\rho}$ and $v_{\rm t} \equiv
\sqrt{\mu / \rho}$ are the longitudinal and transverse bulk sound velocities,
determined by the Lam\'{e} coefficients $\lambda$ and $\mu$. The solution of 
the set of second-order equations of the form (\ref{elasticity equation}), 
or its generalization to anisotropic media, requires boundary conditions on
${\bf u}$ and $({\bf n} \cdot \mbox{\boldmath $\nabla$}) {\bf u}$, where ${\bf 
n}$ is a unit vector normal to the interface.

The conventional boundary conditions applied in this situation (assuming fully
bonded materials)  are as follows\cite{Ewing etal,Royer and Dieulesaint}: 
First, the displacement field is assumed to be continuous across an interface,
\begin{equation}
{\bf u}_{\rm A} = {\bf u}_{\rm B}.
\label{first BC}
\end{equation}
The condition (\ref{first BC}) implies that the two materials are attached and
do not separate. The second condition follows from momentum conservation and
requires that the force density be continuous,
\begin{equation}
T_{\rm A}^{ij} \, n^j = T_{\rm B}^{ij} \, n^j.
\label{second BC} 
\end{equation}
Here $T^{ij}$ is the stress tensor, defined by the continuity equation
\begin{equation}
\partial_t \Pi^i + \partial_j T^{ij} = 0
\end{equation}
for momentum density ${\bf \Pi} \equiv \rho \, \partial_t {\bf u}$, and 
${\bf n}$ is the unit normal\cite{stress tensor footnote}. In an isotropic 
elastic medium, it follows from Eq.~(\ref{elasticity equation}) that the
stress tensor is given by
\begin{eqnarray}
T^{ij} &=& - \lambda (\mbox{\boldmath $\nabla$} \cdot {\bf u}) \, \delta_{ij}
- 2 \mu \, u_{ij} \label{isotropic stress tensor} \\
&=& - \, c_{ijkl} \, u_{kl} ,
\end{eqnarray}
where 
\begin{equation}
c_{ijkl} = \lambda \, \delta_{ij} \delta_{kl} + \mu (\delta_{ik} \delta_{jl}
+ \delta_{il} \delta_{jk}) 
\end{equation} 
is the elastic tensor for a linear isotropic solid, and where
\begin{equation}
u_{ij} \equiv (\partial_i u_j + \partial_j u_i)/2
\label{strain tensor}
\end{equation}
is the strain tensor.

The purpose of this paper is to point out that these boundary conditions,
(\ref{first BC}) and (\ref{second BC}), while quite appropriate for the 
geophysical application mentioned above, are generally incorrect when applied 
to long-wavelength vibrational dynamics in crystals with abrupt, epitaxial 
interfaces. The reason is because in the latter application, elasticity theory
is only an approximate long-wavelength description for the underlying 
microscopic lattice dynamics---which necessarily depends on the detailed 
atomic structure of the interface---whereas (\ref{first BC}) and (\ref{second 
BC}) make no reference to that microscopic structure. For example, the correct
boundary conditions must depend on the effective force constants between type
A and B atoms in Fig.~\ref{superlattice figure}, as well as between atoms of 
the same type.

There are numerous applications of elasticity theory to solid state systems
with heterostructures, where the use of the conventional boundary conditions 
would lead to quantitatively incorrect results. Examples include phonons in
nanostructures such as quantum dots\cite{Leburton etal}, quantum 
wells\cite{Ridley}, superlattices\cite{Camley etal,Tamura etal}, and 
nanoparticles embedded in host materials\cite{Tanaka etal,Ovsyuk and 
Novikov,Zhao and Masumoto}. A correct use of boundary conditions might be 
especially important for nanometer-scale elastic media such as {\it phononic} 
band-gap materials\cite{Garcia-Pablos etal}. Also, the thermal resistance of a
heterojunction is determined by phonon scattering at the interface and
is therefore sensitive to the connection rules or S matrix\cite{Little}.

Finally, we would like to point out a strong analogy between this work and
the problem of determining the appropriate interface boundary conditions for 
the envelope functions in effective mass theory\cite{Bastard review}. In this
case, effective mass theory serves as the appropriate long-wavelength 
approximation to the full Schr\"odinger equation that contains the microscopic
periodic potential of the crystalline lattice, and connection rules are 
required to join envelope functions through an interface between crystals
with different effective mass. The microscopic theory of these connection
rules was first developed by Kroemer and Zhu\cite{Kroemer and Zhu,Zhu and 
Kroemer}, and our work may be regarded as an elasticity theory analog of
Refs. \onlinecite{Kroemer and Zhu} and \onlinecite{Zhu and Kroemer}.

In the next section we give a detailed derivation of the connection rules for
the case of a simple one-dimensional model of an inhomogeneous crystal, a 
chain of harmonic oscillators with an abrupt change in mass and spring 
stiffness parameters, and in Section \ref{numerical studies section} we compare
the results of using both our new connection rules and the conventional 
connection rules to exact results obtained by numerical diagonalization. In 
Section \ref{s matrix section} we relate the connection rule problem to that
of calculating the $S$ matrix for plane-wave scattering from the interface.
The problem of determining the interface boundary conditions between 
three-dimensional solids is discussed in Section \ref{beyond one dimension 
section}, and our conclusions are summarized in Section \ref{discussion 
section}.

\section{connection rules in one dimension}
\label{connection rules section}

We turn now to an analysis of the one-dimensional case, where a chain of atoms
with nearest-neighbor bonds are constrained to move on a line. The vibrations
in this case are purely longitudinal.

\begin{figure}
\centerline{\psfig{file=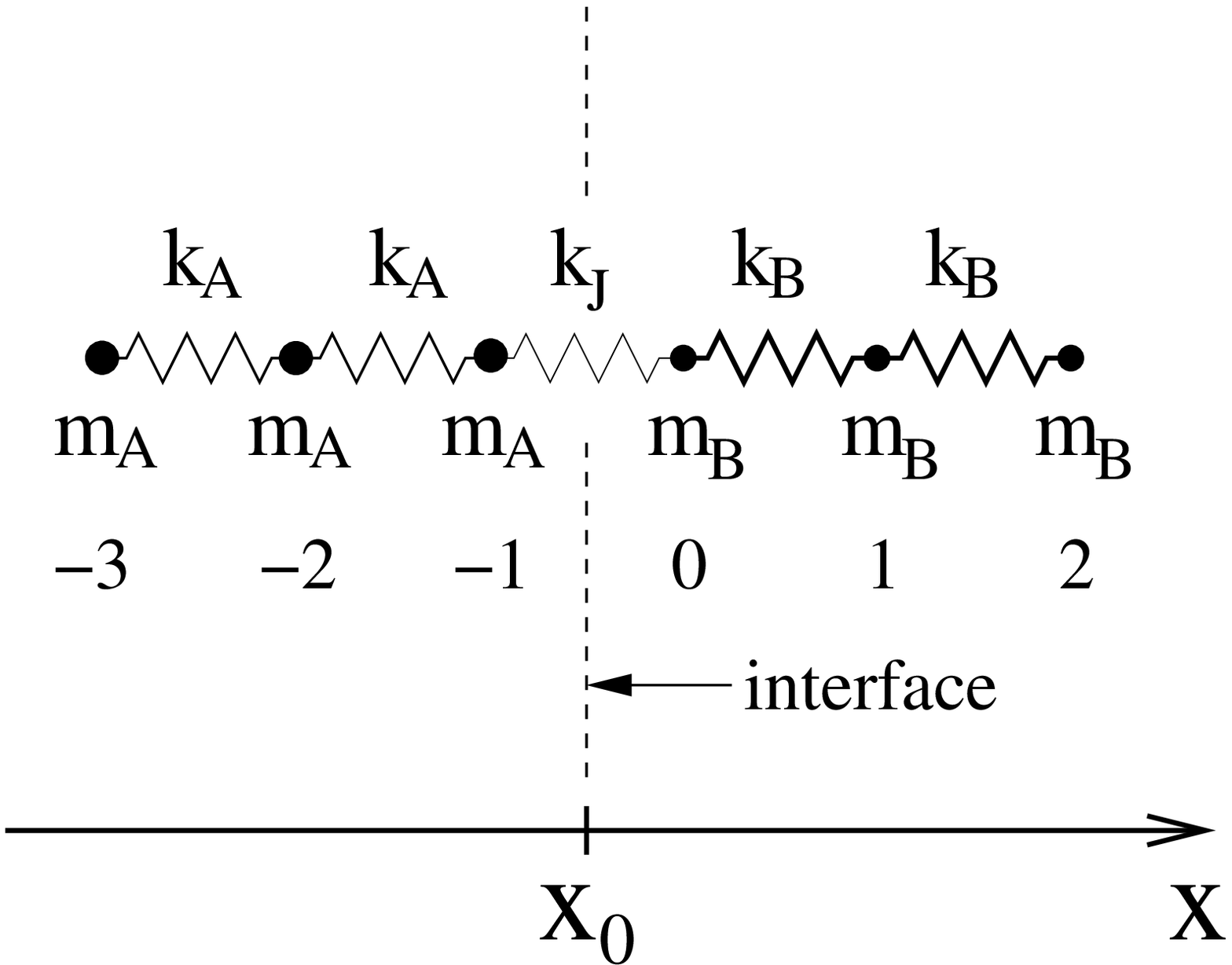,width=2.5in}}
\vspace{0.1in}\setlength{\columnwidth}{3.2in}
\centerline{\caption{
Model of an atomically sharp interface in a one-dimensional crystal.
\label{atomchain figure}}}
\end{figure}

An abrupt interface is introduced at position $x_0$. To the left of $x_0$
the mass of each atom is $m_{\rm A}$, and the effective spring constant of the
nearest-neighbor bonds is $k_{\rm A}$; the corresponding parameters on the 
right side are $m_{\rm B}$ and $k_{\rm B}$. The strength of the bond 
connecting the type A and B atoms, which is generally different from $k_{\rm 
A}$ and $k_{\rm B}$, is denoted by $k_{\rm J}$. The lattice constant on both 
sides is equal to $a$. The model we consider is illustrated in 
Fig.~\ref{atomchain figure}.

According to elasticity theory, which is valid for vibrational wavelengths 
large compared with $a$, the regions to the left and right of the interface 
are described by the wave equations 
\begin{equation}
(\partial_t^2 - v_{\rm I}^2 \partial_x^2) \, u_{\rm I} = 0, \ \ \ \ 
v_{\rm I} \equiv a \sqrt{k_{\rm I}/m_{\rm I}}, \ \ \ \ I=A,B.
\label{wave equations} 
\end{equation}
The elasticity theory description of the homogeneous chain is reviewed in the 
appendix. To proceed, the wave equations (\ref{wave equations}) must be 
supplemented with boundary conditions on $u(x_0)$ and $u'(x_0)$.

A general linear interface boundary condition may be expressed in the form
\begin{equation}
\left[ \matrix{u(x_0) \cr u'(x_0)} \right]_{\rm B} = M
\left[ \matrix{u(x_0) \cr u'(x_0)} \right]_{\rm A},
\label{M definition} 
\end{equation}
where $M$ is a $2 \times 2$ matrix. The connection rule matrix implied by the
boundary conditions (\ref{first BC}) and (\ref{second BC}) is
\begin{equation}
M = \left(\matrix{ 1 & 0 \cr 0 & k_{\rm A}/k_{\rm B} }\right).
\label{old M}
\end{equation}
A common application of (\ref{old M}) is to an elastic string with an abrupt 
change in mass density, but no change in elasticity\cite{French,Fetter and 
Walecka}; in this case (\ref{old M}) reduces to the identity 
matrix\cite{string footnote}.

It is simple to demonstrate that (\ref{old M}) is the only matrix consistent 
with conditions (\ref{first BC}) and (\ref{second BC}): First, continuity
requires that $M_{11}=1$ and $M_{12}=0.$ To find the other elements, we note 
that in one dimension the $xx$ component of the stress tensor (\ref{isotropic 
stress tensor}) is $T^{xx} = - \rho v^2 \partial_x u$. The stress immediately 
to the left of the interface is therefore $T^{xx}_{\rm A} = -k_{\rm A} 
u'_{\rm A}(x_0)$, and that to the immediate right is $T^{xx}_{\rm B} = -
k_{\rm B} u'_{\rm B}(x_0)$. Now, Eq.~(\ref{M definition}) requires that
\begin{equation}
k_{\rm B} \, u'_{\rm B}(x_0) = k_{\rm B} \bigg( M_{21} \, u_{\rm A}(x_0) +  
M_{22} \, u'_{\rm A}(x_0) \bigg),
\end{equation}
which implies
\begin{equation}
T^{xx}_{\rm B} = - M_{21} \, k_{\rm B} \, u_{\rm A}(x_0) + M_{22} \
(k_{\rm B}/k_{\rm A}) \ T^{xx}_{\rm A}.
\label{stress condition}
\end{equation}
Therefore, the condition (\ref{second BC}) requires that $M_{21}=0$ and
$M_{22} = k_{\rm A}/k_{\rm B}$.

We now proceed with our derivation of the correct boundary condition matrix 
$M$ for the model shown in Fig.~\ref{atomchain figure}. The coordinates of 
the atoms are written as
\begin{equation}
x_n(t) = x_n^0 + \xi_n(t) , \ \ \ \ \ \ x_n^0 \equiv na.
\end{equation}
The equation of motion for atom $n$ is
\begin{equation}
m_n \, \ddot{\xi}_n = k_{\rm r} \, (\xi_{n+1}-\xi_n) - k_{\rm l} \, (\xi_n - 
\xi_{n-1}), 
\label{general equation of motion}
\end{equation}
where $k_{\rm r}$ is the stiffness of the spring to the right of mass $m_n$, 
and $k_{\rm l}$ is that to the left. Assuming harmonic time dependence we have,
for the atoms immediately to the left ($n=-1$) and right ($n=0$) of the 
interface,
\begin{equation}
- \omega^2 \, m_{\rm A} \, \xi_{-1} = k_{\rm J} (\xi_{0} - \xi_{-1}) - 
k_{\rm A} (\xi_{-1} - \xi_{-2})
\label{n=-1 equation}
\end{equation}
and
\begin{equation}
- \omega^2 \, m_{\rm B} \, \xi_{0} = k_{\rm B} (\xi_{1} - \xi_{0}) - k_{\rm J}
(\xi_{0} - \xi_{-1}).
\label{n=0 equation}
\end{equation}
Next we introduce the displacement field $u(x)$ as a smooth interpolating 
function between the $\xi_n$, such that
\begin{equation}
u(x^0_n) = \xi_n,
\label{displacement field definition}
\end{equation}
and use the following relations,
\begin{eqnarray}
\xi_{-2} &=& u_{\rm A}(x_0 - {\textstyle{3 \over 2}}a) \approx u_{\rm A}(x_0) -
{\textstyle{3 \over 2}}a \, u_{\rm A}'(x_0), \\
\xi_{-1} &=& u_{\rm A}(x_0 - {\textstyle{1 \over 2}}a) \approx u_{\rm A}(x_0) -
{\textstyle{1 \over 2}}a \, u_{\rm A}'(x_0), \\
\xi_{0} &=& u_{\rm B}(x_0 + {\textstyle{1 \over 2}}a) \approx u_{\rm B}(x_0) +
{\textstyle{1 \over 2}}a \, u_{\rm B}'(x_0), \\
\xi_{1} &=& u_{\rm B}(x_0 + {\textstyle{3 \over 2}}a) \approx u_{\rm B}(x_0) +
{\textstyle{3 \over 2}}a \, u_{\rm B}'(x_0). 
\end{eqnarray}
Because the interface boundary conditions involve the displacement field and 
its first derivative only, second and higher-order gradients are neglected 
here. Furthermore, as the frequency $\omega$ is formally of the order of a 
gradient 
(recall the bulk dispersion relation $\omega = v |k|$), for consistency we 
also neglect the terms proportional to $\omega^2$ in Eqs.~(\ref{n=-1 
equation}) and (\ref{n=0 equation})\cite{frequency foonote}.

The resulting coupled equations can be put in the form
\begin{eqnarray}
&\left( \matrix{ k_{\rm J} & {1 \over 2} a k_{\rm J} \cr -k_{\rm J} &
a(k_{\rm B}-{1 \over 2} k_{\rm J}) \cr }\right)&
\left[ \matrix{u(x_0) \cr u'(x_0)} \right]_{\rm B} \nonumber \\
= &\left( \matrix{ k_{\rm J} & a (k_{\rm A} - {1 \over 2} k_{\rm J}) 
\cr -k_{\rm J} & {1 \over 2} a k_{\rm J} \cr }\right)&
\left[ \matrix{u(x_0) \cr u'(x_0)} \right]_{\rm A},
\end{eqnarray}
which, upon comparison with (\ref{M definition}), identifies
\begin{equation}
\left( \matrix{ k_{\rm J} & {1 \over 2} a k_{\rm J} \cr -k_{\rm J} & 
a(k_{\rm B}-{1 \over 2} k_{\rm J}) \cr }\right)^{-1}
\left( \matrix{ k_{\rm J} & a (k_{\rm A} - {1 \over 2} k_{\rm J}) 
\cr -k_{\rm J} & {1 \over 2} a k_{\rm J} \cr }\right)
\end{equation}
as the connection rule matrix. Therefore we obtain, for the model shown in
Fig.~\ref{atomchain figure}, the connection rules
\begin{equation}
M = \left( \matrix{ 1 & a[k_{\rm A}k_{\rm B}-{1\over 2} k_{\rm J}(k_{\rm A}+ 
k_{\rm B})] /k_{\rm J} k_{\rm B} \cr 0 & k_{\rm A}/k_{\rm B} \cr }\right).
\label{new M}
\end{equation}

Several remarks are in order: First, the correct connection rules clearly 
depend on the microscopic structure of the interface, including the stiffness
$k_{\rm J}$ of the interface bond, which is generally different than
$k_{\rm A}$ and $k_{\rm B}$. The boundary conditions cannot be deduced by
conservation laws that do not make reference to the microscopic structure.
Second, the matrix (\ref{new M}) is generally
off-diagonal, implying a connection between the displacement field $u$ on one 
side of the interface, with the strain $u'$, as well as the 
displacement, on the other. Third, the displacement field is generally {\it 
not} continuous through the interface, in contrast with the conventional
assumption. This discontinuity, however, does not imply that the two sides are
separated. It simply means that the atomic displacements $\xi_n$, when
extrapolated from each side to the ``mathematical interface'' at $x_0$, do not
meet. Fourth, we note that in the limit $a \rightarrow 0$ the boundary 
conditions (\ref{old M}) and (\ref{new M}) agree. However, this limit is not 
meaningful in a real crystal. And finally, we note that (\ref{old M}) and 
(\ref{new M}) also become equivalent in the event that $k_{\rm J}$ has the 
special value $k^*_{\rm J}$ given by
\begin{equation}
{1 \over k^*_{\rm J}} = {1 \over 2}\bigg( {1 \over k_{\rm A}} + {1 \over 
k_{\rm B}} \bigg).
\end{equation} 

\section{numerical studies}
\label{numerical studies section}

When $k_{\rm J}$ differs from $k^*_{\rm J}$, the influence of the off-diagonal
element in (\ref{new M}) can become substantial. To demonstrate this we use
elasticity theory with (\ref{old M}) and (\ref{new M}) to predict the normal 
modes frequencies of a one-dimensional inhomogeneous crystal of finite length 
$L$, and compare both with the exact spectrum obtained numerically. The
interface is placed at $x_0 = L/2$.

The elasticity theory spectrum is obtained by (numerically) searching for 
frequencies such that the three conditions $u(0)=0$, $u(L)=0$, and (\ref{M 
definition}) are satisfied. The appropriate solution of the wave equation
to the left of the interface, on the interval $0 \le x \le x_0$, is 
\begin{equation}
u_{\rm A}(x) = \sin(\omega x/v_{\rm A}),
\label{left solution}
\end{equation}
and to the right ($x_0 \le x \le L$) is
\begin{equation}
u_{\rm B}(x) = \alpha \cos(\omega x/v_{\rm B}) + \beta \sin(\omega x/v_{\rm 
B}).
\label{right solution}
\end{equation}
$\alpha$ and $\beta$ are uniquely determined (at each frequency) by the 
requirement that (\ref{M definition}) is satisfied. This leads to 
\begin{equation}
\left[ \matrix{u(x_0) \cr u'(x_0)} \right]_{\rm B} = C  \left[ \matrix{ 
\alpha \cr \beta} \right] = M  \left[ \matrix{u(x_0) \cr u'(x_0)} 
\right]_{\rm A},
\label{interface condition}
\end{equation}
where 
\begin{equation}
C \equiv \left( \matrix{ \cos(\omega L/2 v_{\rm B}) & \sin(\omega L/2 
v_{\rm B}) \cr -(\omega/v_{\rm B}) \, \sin(\omega L/2 v_{\rm B}) & 
(\omega/v_{\rm B}) \, \cos(\omega L/2 v_{\rm B})} \right).
\label{C matrix}
\end{equation}
From (\ref{interface condition}) we obtain $\alpha(\omega)$ and 
$\beta(\omega)$ as
\begin{equation}
\left[ \matrix{ \alpha \cr \beta} \right] = C^{-1} M 
\left[ \matrix{ \sin(\omega L / 2 v_{\rm A}) \cr (\omega / v_{\rm A}) \,
\cos(\omega L / 2 v_{\rm A}) } \right],
\end{equation}
and the normal mode frequencies from the remaining boundary condition 
$u_{\rm B}(L)=0$.

The exact spectrum is obtained by expressing the coupled equations of motion 
(\ref{general equation of motion}) for a chain of $N$ atoms, with the first 
and last atoms held fixed, as a nonsymmetric eigenvalue problem. The system 
size is then given by $L = N a$. For the results presented below, $N = 101.$

\begin{figure}
\centerline{\psfig{file=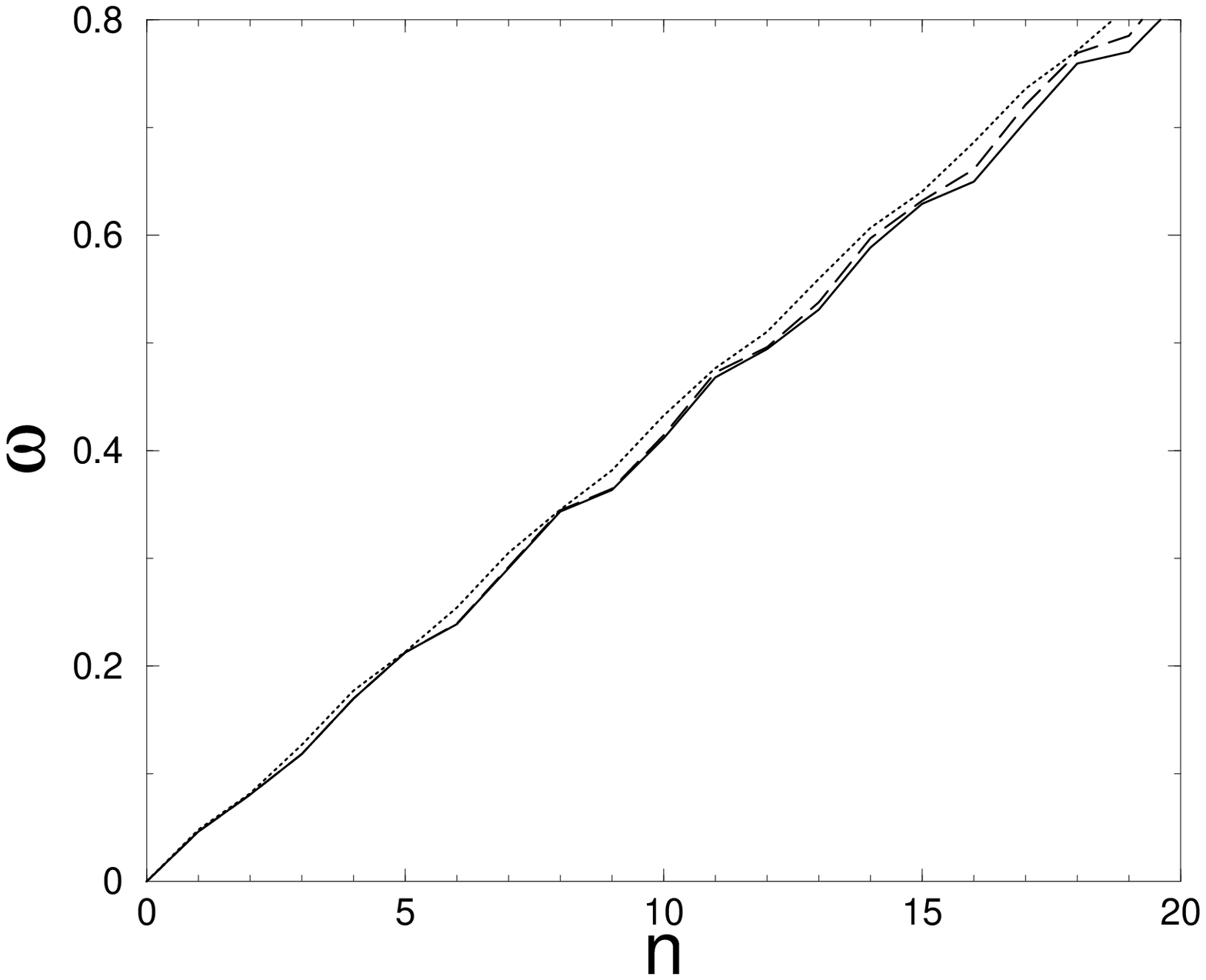,width=3.0in}}
\vspace{0.1in}\setlength{\columnwidth}{3.2in}
\centerline{\caption{
Vibrational spectrum with $k_{\rm B} / k_{\rm A} = 5.0$ and $k_{\rm J} / 
k_{\rm A} = 0.20$. 
\label{case2 figure}}}
\end{figure}

\begin{figure}
\centerline{\psfig{file=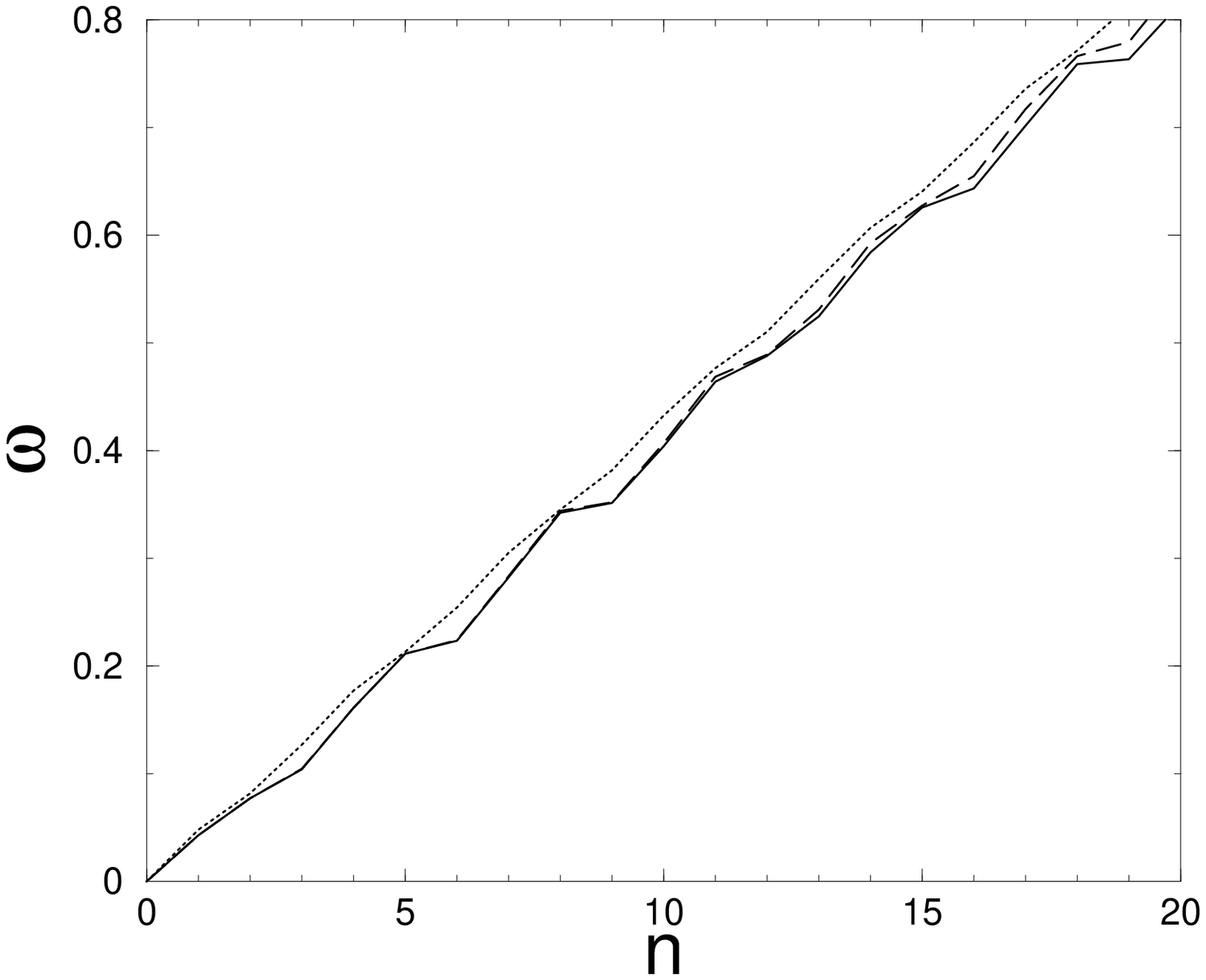,width=3.0in}}
\vspace{0.1in}\setlength{\columnwidth}{3.2in}
\centerline{\caption{
Vibrational spectrum with $k_{\rm B} / k_{\rm A} = 5.0$ and $k_{\rm J} / 
k_{\rm A} = 0.05$. 
\label{case4 figure}}}
\end{figure}

\begin{figure}
\centerline{\psfig{file=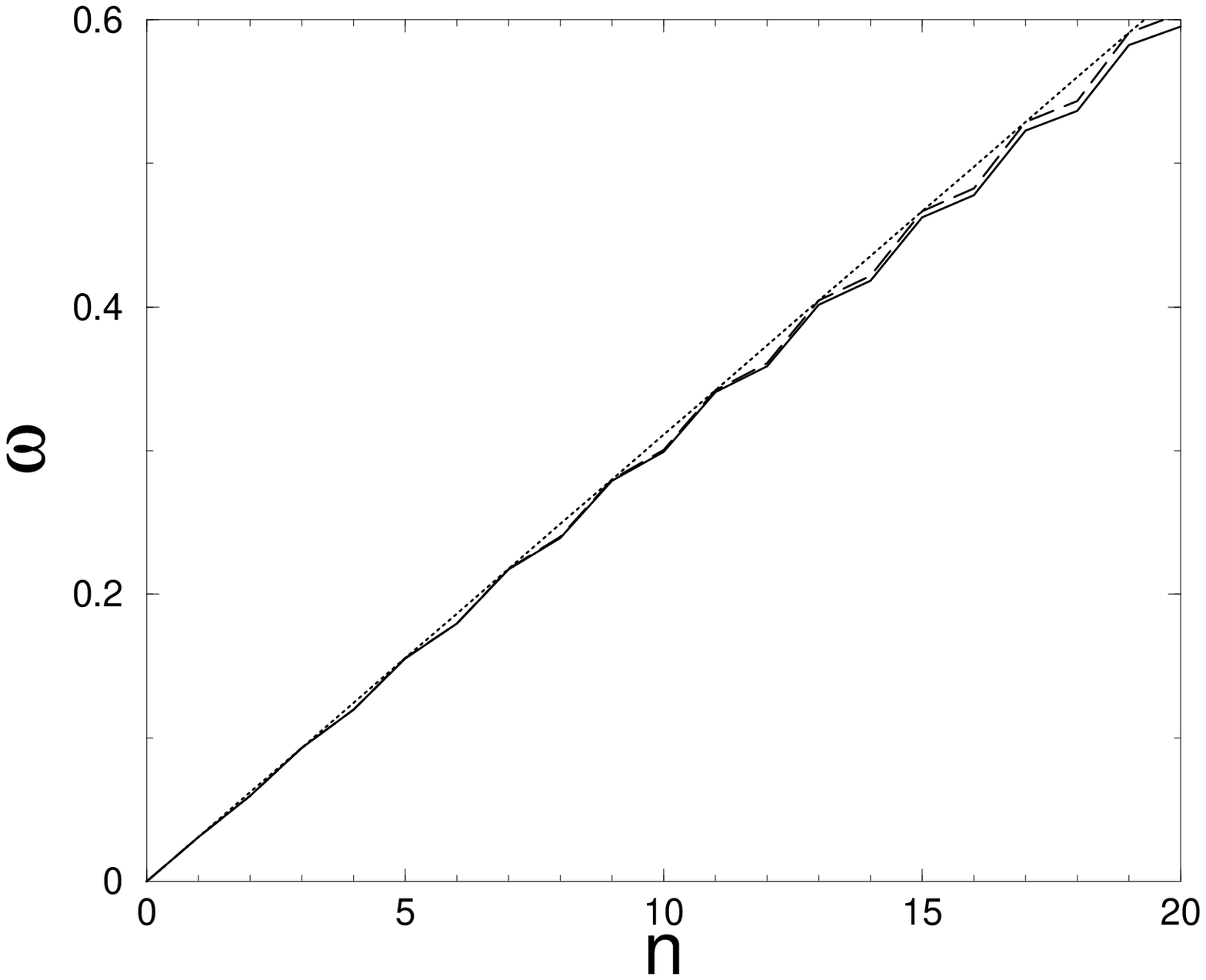,width=3.0in}}
\vspace{0.1in}\setlength{\columnwidth}{3.2in}
\centerline{\caption{
Vibrational spectrum with $k_{\rm B} / k_{\rm A} = 1.0 $ and $k_{\rm J} / 
k_{\rm A} = 0.20$. 
\label{case3 figure}}}
\end{figure}

Representative results are shown in Figs.~\ref{case2 figure} through 
\ref{case3 figure}. In each case the frequency $\omega$ of mode $n$ is given 
in units of $\pi v_{\rm A} / L$. Figs.~\ref{case2 figure} and \ref{case4 
figure} the show vibrational spectra of two inhomogeneous chains, both with 
$k_{\rm B} = 5.0 \, k_{\rm A}.$ The curves in these figures are independent of
the masses $m_{\rm A}$ and $m_{\rm B}$; the only mass dependence is in the 
energy scale $\pi v_{\rm A} / L.$  In each case the solid line is the exact 
spectrum, the dotted line is the elasticity theory spectrum calculated with 
the conventional connection rules (\ref{old M}), and the dashed line is the 
elasticity theory spectrum calculated with the connection rules (\ref{new M}).
In Fig.~\ref{case2 figure}, $k_{\rm J} = 0.20 \, k_{\rm A}$, and the three 
spectra are similar. In Fig.~\ref{case4 figure}, where $k_{\rm J} = 0.05 \, 
k_{\rm A}$, the two sides are only weakly bonded together, and the spectrum 
calculated with Eq.~(\ref{new M}) agrees with the exact spectrum, whereas the 
spectrum calculated with Eq.~(\ref{old M}) does not. At higher frequencies 
both elasticity theory spectra deviate from the exact spectrum because the 
wavelengths become shorter.

The final set of spectra we present, shown in Fig.~\ref{case3 figure}, 
corresponds to a homogeneous chain, $k_{\rm B} = k_{\rm A},$ with a weakly 
bonded interface, $k_{\rm J} = 0.20 \, k_{\rm A}$. The spectrum calculated 
with Eq.~(\ref{new M}) agrees well with the exact spectrum. The elasticity 
theory spectrum calculated with Eq.~(\ref{old M}) misses the fine structure 
present in the exact spectrum because Eq.~(\ref{old M}) makes no reference to 
the value of $k_{\rm J}$.

\section{S matrix}
\label{s matrix section}

\begin{figure}
\centerline{\psfig{file=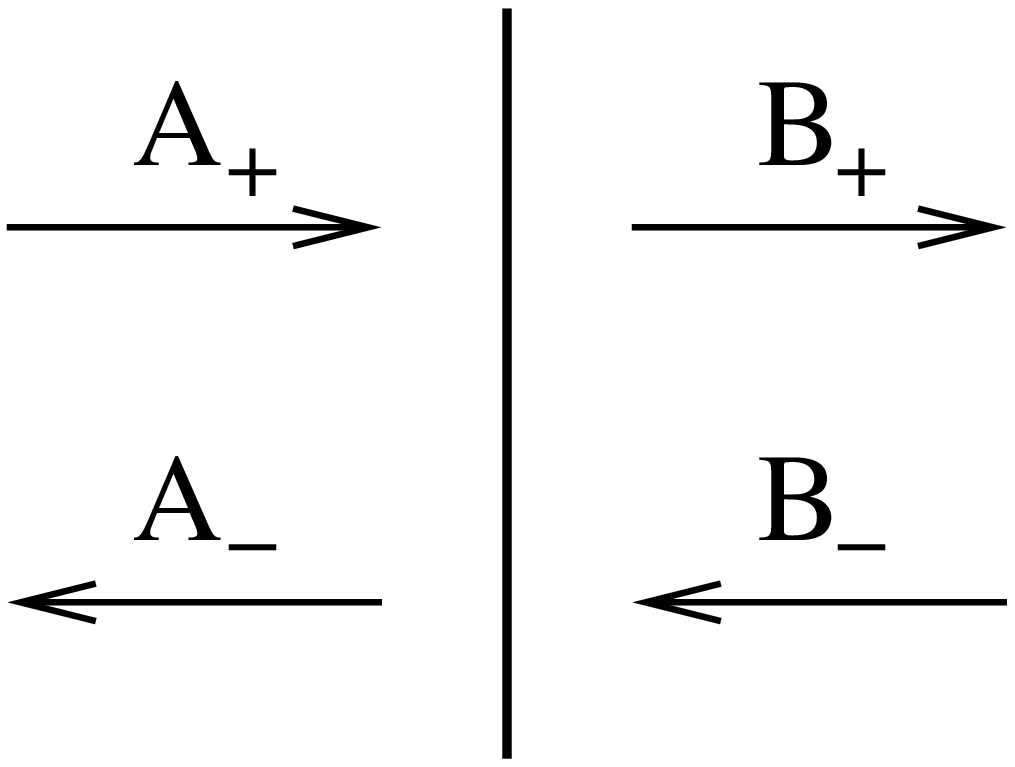,width=2.0in}}
\vspace{0.1in}\setlength{\columnwidth}{3.2in}
\centerline{\caption{
Incoming and outgoing waves related by the S matrix. The interface is at 
$x = 0$.
\label{s matrix figure}}}
\end{figure}

An alternative but physically equivalent way of expressing the interface 
boundary conditions is through an S matrix. Whereas the matrix $M$ gives
the linear relation between the displacement field $u(x_0)$ and
its derivative $u'(x_0)$ on side A to that on side B, the S matrix
relates the amplitudes of waves incident on the interface, from both sides,
to the corresponding outgoing waves. In this case we take $x_0$ to be at the
origin and we write the elasticity theory solutions as\cite{planewave footnote}
\begin{equation}
u_{\rm A}(x) = A_{+} \, e^{i \omega x/v_{\rm A}} + A_{-} \,  e^{-i \omega 
x/v_{\rm A}} 
\label{left S expansion}
\end{equation}
and
\begin{equation}
u_{\rm B}(x) = B_{+} \, e^{i \omega x/v_{\rm B}} + B_{-} \,  e^{-i \omega 
x/v_{\rm B}},
\label{right S expansion}
\end{equation}
where $A_{\pm}$ and $B_{\pm}$ are complex coefficients giving the amplitudes
of the plane waves shown in Fig.~\ref{s matrix figure}.

The S matrix relates the coefficients in Eqs.~(\ref{left S expansion}) and
(\ref{right S expansion}), and is defined by
\begin{equation}
\left[ \matrix{A_{-} \cr B_{+}} \right] = S
\left[ \matrix{A_{+} \cr B_{-}} \right].
\label{S definition} 
\end{equation}
From (\ref{M definition}) we obtain
\begin{equation}
\left[ \matrix{B_{+} \cr B_{-}} \right] = {\cal M} \left[ \matrix{A_{+} 
\cr A_{-}} \right]
\label{S relation} 
\end{equation}
and therefore
\begin{equation}
S = {1 \over {\cal M}_{22}} \left( \matrix{ -{\cal M}_{21} & 1  \cr 
{\rm Det} \, {\cal M}    &  {\cal M}_{12} } 
\right),
\label{S matrix}
\end{equation}
where
\begin{equation}
{\cal M} \equiv \left( \matrix{ 1 & 1  \cr i \omega/v_{\rm B}  & -i \omega
/ v_{\rm B}} \right)^{-1} \! \! \! M \left( \matrix{ 1 & 1  \cr i \omega /
v_{\rm A}  & -i \omega / v_{\rm A}} \right).
\label{cal M} 
\end{equation}
Here ${\rm Det} {\cal M}$ is the determinant of ${\cal M}$. A useful
expression for ${\cal M}$ may be obtained by combining Eqs.~(\ref{old M})
and (\ref{new M}) as
\begin{equation}
M = \left(\matrix{ 1 & M_{12} \cr 0 & k_{\rm A}/k_{\rm B} }\right),
\label{combined M}
\end{equation}
where $M_{12}$ is either equal to zero or to the off-diagonal element in 
(\ref{new M}). Using this representation for $M$ we find
\begin{equation}
{\cal M} = {1 \over 2} \left( \matrix{ 
1 + {k_{\rm A} v_{\rm B} \over k_{\rm B} v_{\rm A}} + i  M_{12} {\omega \over
v_{\rm A}}  & 
1 - {k_{\rm A} v_{\rm B} \over k_{\rm B} v_{\rm A}} - i  M_{12} {\omega \over
v_{\rm A}} \cr 
1 - {k_{\rm A} v_{\rm B} \over k_{\rm B} v_{\rm A}} + i M_{12} {\omega \over 
v_{\rm A}}   & 
1 + {k_{\rm A} v_{\rm B} \over k_{\rm B} v_{\rm A}} - i  M_{12} {\omega \over
v_{\rm A}}  } \right)
\end{equation}
and
\begin{equation}
{\rm Det} \, {\cal M} = k_{\rm A} v_{\rm B} \big/  k_{\rm B} v_{\rm A}. 
\end{equation}
Note that the complex terms in the S matrix come from the off-diagonal element
in (\ref{new M}). 

The S matrix provides a simple and direct way to obtain transmission and 
reflection amplitudes, ${\sf t}$ and ${\sf r}$, for scattering from the 
interface. From 
(\ref{S matrix}) we observe that the transmission and reflection amplitudes 
for a wave of unit amplitude incident from the left $(A_+ = 1 \ \ {\rm and} 
\ \ B_- = 0)$ are
\begin{equation}
{\sf t} = {{\rm Det} \, {\cal M} \over {\cal M}_{22}} = {2 \, k_{\rm A} 
v_{\rm B} \over {k_{\rm A} v_{\rm B} +  k_{\rm B} v_{\rm A}  - i  M_{12} \, 
\omega \, k_{\rm B}}}
\label{transmission coefficient}
\end{equation}
and
\begin{equation}
{\sf r} = - {{\cal M}_{21} \over {\cal M}_{22}} = {k_{\rm A} v_{\rm B} -  
k_{\rm B} v_{\rm A}  - i  M_{12} \, \omega \, k_{\rm B} \over k_{\rm A} 
v_{\rm B} +  k_{\rm B} v_{\rm A}  - i  M_{12} \, \omega \, k_{\rm B}}.
\label{reflection coefficient}
\end{equation}
In the limit $k_{\rm A} = k_{\rm B} = k_{\rm J}$ where the mass density is
discontinuous but the elasticity is continuous, these amplitudes reduce to
\begin{equation}
{\sf t} \rightarrow {2 v_{\rm B} \over  v_{\rm B} +  v_{\rm A}} \ \ \ \ \ 
{\rm and} \ \ \ \ \ {\sf r} \rightarrow {v_{\rm B} -  v_{\rm A} \over  
v_{\rm B} +  v_{\rm A}},
\end{equation}
the well-known results for scattering from a mass discontinuity\cite{Fetter
and Walecka}. It can be shown that the transmission and reflection 
coefficients, ${\sf T}$ and ${\sf R}$, defined as the fraction of transmitted 
and reflected energy flux, are determined from Eqs.~(\ref{transmission 
coefficient}) and (\ref{reflection coefficient}) according to
\begin{equation}
{\sf T} = {v_{\rm A} k_{\rm B}\over v_{\rm B} k_{\rm A}} \, \big| {\sf t} 
\big|^2 \ \ \ \ \ {\rm and} \ \ \ \ \ {\sf R} =  \big| {\sf r} \big|^2. 
\end{equation}

In addition to relating the connection rule matrix $M$ to observable 
quantities, this scattering theory formulation serves to reemphasize the main 
thesis of this paper, that the connection rules must depend on the microscopic
structure of the heterojunction and cannot be determined by ``far field'' 
information alone.

\section{beyond one dimension}
\label{beyond one dimension section}

In this section we give a brief discussion of the generalization of our method
to three-dimensional epitaxial heterojunctions. To allow for both longitudinal
and transverse elastic waves one must work with a $6 \times 6$ connection
matrix ${\tilde M}$ satisfying
\begin{equation}
\left[ \matrix{u_x(x_0) \cr u_y(x_0) \cr u_z(x_0) \cr u'_x(x_0) \cr  u'_y(x_0)
\cr u'_z(x_0)} \right]_{\rm B} = {\tilde M}
\left[ \matrix{u_x(x_0) \cr u_y(x_0) \cr u_z(x_0) \cr u'_x(x_0) \cr  u'_y(x_0)
\cr u'_z(x_0)} \right]_{\rm A}.
\label{M tilde definition} 
\end{equation}
Here $ u'_i \equiv {\bf n} \cdot \mbox{\boldmath $\nabla$} u_i$, with ${\bf n}$
a unit vector normal to the interface, and $i = x, y, z.$ The procedure for 
obtaining ${\tilde M}$ is identical to that described in Section 
\ref{connection rules section}; however, in general it will be necessary to 
include atomic bonds beyond those connecting nearest-neighbor atoms.

To obtain quantitatively accurate connection rules one would need to determine
the atomic structure of the particular interface and the required force 
constants. This can be accomplished using first-principles electronic structure
calculation methods (for example, those based on density functional theory),
although a full treatment of a three-dimensional heterojunction would be very
demanding computationally.

\section{discussion}
\label{discussion section}

We have shown that the conventional interface boundary conditions used in 
elasticity theory, requiring that the displacement field and its associated 
stress field be continuous, are generally incorrect for epitaxial interfaces. 
The correct boundary conditions are nonuniversal and depend on the detailed 
microscopic structure of the heterojunction. 

The conventional boundary conditions are incorrect because the displacement
field ${\bf u}({\bf r})$ is generally discontinuous. However, this 
discontinuity does {\it not} imply that the two sides separate. In the
elasticity theory description of crystalline lattice dynamics, 
\begin{equation}
{\bf u}({\bf r}_0) = {\bf r}_n - {\bf r}^0_n
\end{equation}
is simply a function giving the displacement of atom $n$ at each equilibrium
lattice point ${\bf r}_n^0$. A discontinuity in ${\bf u}({\bf r})$ at a 
``mathematical'' interface between layers of atoms implies that the atomic 
displacements ${\bf r}_n - {\bf r}^0_n $ on each side of an  
interface do not meet when smoothly interpolated to that interface. In 
contrast, the condition that the stress be continuous follows from momentum 
conservation and is generally correct\cite{momentum footnote}.

It is tempting to approach the interface boundary condition problem by using
elasticity equations generalized to the case of a compositionally graded 
crystal, characterized by a position-dependent mass density and elastic
parameters, and then take the limit of an abrupt composition change. But this
too is incorrect, for elasticity theory is intrinsically a long-wavelength 
description and can be formulated only for slowly graded systems, making
the required limit invalid. 

For example, the generalized wave equation describing the long-wavelength
vibrational dynamics in a one-dimensional crystal with lattice constant $a$,
mass density $\rho(x)$, and stiffness $k(x)$, can be shown to be (see appendix)
\begin{equation}
[\rho(x) \partial_t^2 - a \partial_x k(x) \partial_x] u(x,t) = 0.
\label{graded wave equation}
\end{equation}
Integration of (\ref{graded wave equation}) shows that $u(x)$ and $k(x) \, 
u'(x)$ are continuous, consistent with the conventional boundary conditions 
(\ref{old M}). However, Eq.~(\ref{graded wave equation}), which neglects
stiffness gradients higher order than $k'(x)$, is not valid in the 
abrupt limit.

Having made the case that the conventional interface boundary conditions
(\ref{first BC}) and (\ref{second BC}) do not apply to epitaxial interfaces,
we must emphasize again that we have not provided generally applicable
conditions to replace (\ref{first BC}) and (\ref{second BC}). The connection
rules (\ref{new M}) are only valid for the simple one-dimensional interface
model shown in Fig.~\ref{atomchain figure}.

In closing, we would like to speculate about the reason the subject of this
paper has been, to the best of our knowledge, overlooked in the solid state
physics literature. Historically, elasticity theory was developed as a 
self-contained branch of mechanics that made no reference to a possible 
underlying atomic structure, and much of the theory was developed before the 
wide acceptance of the atomic view of matter. The conventional boundary 
conditions (\ref{first BC}) and (\ref{second BC}) are certainly correct within
elasticity theory proper. However, within solid state physics, elasticity 
theory is regarded as a long-wavelength description with a well-defined but
limited regime of validity, and we believe that the connection rules in 
question were applied to heterostructures without considering that regime of
validity.

\acknowledgements

This work was supported by the Research Corporation. It is a pleasure to thank
Steve Lewis and Kelly Patton for useful discussions.

\appendix

\section{homogeneous chain}
\label{homognenous chain section}

Here we record the long-wavelength theory of the homogeneous harmonic 
oscillator chain with masses $m$, spring constants $k$, and lattice constant
$a$. In this case the equation of motion leads to
\begin{equation}
\partial_t^2 u(x,t) - {k \over m} \big[ u(x+a,t) - 2 \, u(x,t) + u(x-a,t) \big]
=0.
\label{uniform chain}
\end{equation}
Taylor expanding (\ref{uniform chain}) leads to the one-dimensional wave 
equation
\begin{equation}
(\partial_t^2 - v^2 \partial_x^2) \, u(x,t) = 0,
\label{homogeneous wave equation}
\end{equation}
with sound velocity 
\begin{equation}
v \equiv a \sqrt{k/m}.
\label{homogeneous sound velocity}
\end{equation}

Next we derive the momentum conservation condition satisfied by the 
displacement field $u$. The momentum density carried by a longitudinal elastic
wave in one-dimension is $\Pi = \rho \, \partial_t u,$ where $\rho$ is the 
mass density. In the absence of external forces, Eq.~(\ref{homogeneous wave 
equation}) shows that $\Pi$ satisfies the continuity equation
\begin{equation}
\partial_t \Pi + \partial_x T = 0, 
\label{momentum continuity equation}
\end{equation}
where
\begin{equation}
T = - \rho v^2 \partial_x u
\label{wave equation stress tensor}
\end{equation}
is the scalar stress. As expected, (\ref{wave equation stress tensor}) is 
identical to the $xx$ component of the stress tensor (\ref{isotropic stress 
tensor}).
Similarly, the energy density ${\cal E} = {1 \over 2} \rho [ (\partial_t u)^2
+ v^2 (\partial_x u)^2]$ satisfies the continuity equation
\begin{equation}
\partial_t {\cal E} + \partial_x j_{\rm e} = 0,
\end{equation}
where
\begin{equation}
j_{\rm e} = - \rho v^2 \partial_x u \, \partial_t u
\end{equation}
is the energy flux.

The long-wavelength description of a harmonic oscillator chain with spatially 
varying masses and spring constants follows from the appropriate gradient
expansion of
\begin{eqnarray}
m(x) \, \partial_t^2 u(x,t) &=& k(x+{\textstyle{a \over 2}}) [u(x+a) - u(x)] 
\nonumber \\
&-& k(x-{\textstyle{a \over 2}}) [u(x) - u(x-a)]. 
\end{eqnarray}
Neglecting gradients beyond $k'(x)$ leads to the form (\ref{graded 
wave equation}) quoted in Section\ref{discussion section}.

\end{multicols}

\begin{references}
\bibitem{stress tensor footnote} Note that our definition of the stress tensor
differs by a sign from that often adopted in elasticity theory.
\bibitem{string footnote} Our connection-rule matrix (\ref{new M}) reduces to 
the identity matrix in this case as well.
\bibitem{frequency foonote} It is possible to obtain connection rules without
neglecting the terms proportional to $\omega^2,$ but then the connection rules
would, of course, be frequency dependent.
\bibitem{planewave footnote} The physical displacement fields, which are real,
are given in this case by the real parts of $u_{\rm A}(x) \, e^{-i \omega t}$ 
and $u_{\rm B}(x) \, e^{-i \omega t}$
\bibitem{momentum footnote} Recall from Eq.~(\ref{stress condition}) that
the $M_{21}$ and $M_{22}$ elements are determined by momentum conservation,
whereas $M_{11}$ and $M_{12}$ are determined by displacement field continuity.
\end{references}
\end{document}